\documentstyle[12pt,subeqn]{article}
\newfont{\twelvemsb}{msbm10 scaled\magstep1}
\newfont{\eightmsb}{msbm8}
\newfont{\sixmsb}{msbm6}
\newfam\msbfam
\textfont\msbfam=\twelvemsb
\scriptfont\msbfam=\eightmsb
\scriptscriptfont\msbfam=\sixmsb
\catcode`\@=11
\def\Bbb{\ifmmode\let\next\Bbb@\else
  \def\next{\errmessage{Use \string\Bbb\space only in math mode}}\fi\next}
\def\Bbb@#1{{\Bbb@@{#1}}}
\def\Bbb@@#1{\fam\msbfam#1}
\newfont{\twelvegoth}{eufm10 scaled\magstep1}
\newfont{\tengoth}{eufm10}
\newfont{\eightgoth}{eufm8}
\newfam\gothfam
\textfont\gothfam=\twelvegoth
\scriptfont\gothfam=\eightgoth
\def\frak{\ifmmode\let\next\frak@\else
  \def\next{\errmessage{Use \string\frak\space only in math mode}}\fi\next}
\def\frak@#1{{\fam\gothfam{{#1}}}}
\catcode`@=12
\newcommand{\be}{\begin{equation}}
\newcommand{\ee}{\end{equation}}
\newcommand{\bea}{\begin{eqnarray}}
\newcommand{\ena}{\end{eqnarray}}
\newcommand{\sect}[1]{\setcounter{equation}{0}\section{#1}}

\newcommand{\ZZ}{{\Bbb Z}}
\newcommand{\ZP}{{\ZZ'}}
\renewcommand{\a}{\alpha}
\renewcommand{\b}{\beta}
\newcommand{\e}{\varepsilon}

\newcommand{\fra}{{\frak a}}
\newcommand{\frb}{{\frak b}}

\newcommand{\vs}[1]{\vspace{#1 mm}}
\newcommand{\shalf}{\frac{1}{2}}
\newcommand{\smbox}[1]{\ \mbox{#1}\ }
\newcommand{\medbox}[1]{\quad\mbox{#1}\quad}
\newcommand{\bigbox}[1]{\qquad\mbox{#1}\qquad}
\newcommand{\cA}{{\cal A}}

\newcommand{\vu}{\mbox{\bf u}}
\newcommand{\vx}{\mbox{\bf x}}
\newcommand{\vy}{\mbox{\bf y}}
\newcommand{\ta}{{\tilde a}}
\newcommand{\wA}{{\widehat A}}

\newcommand{\uqa}{U_q({\widehat A}_{N-1})}

\newtheorem{theorem}{Theorem}
\input{epsf}

\begin{document}
\newpage
\pagestyle{empty}
\setcounter{page}{0}
\vfill
\begin{center}

  {\LARGE {\bf {\sf ANYONIC REALIZATION OF THE

                \vs{5}

                QUANTUM AFFINE LIE ALGEBRA $\uqa$}}}\\[1cm]

  \vs{10}

  {\large L. Frappat$^{a}$,
    A. Sciarrino$^{b}$\footnote{supported in part by MURST.},
    S. Sciuto$^{c1}$,
    P. Sorba$^{a}$
  }

  \vs{10}

  {\em $^{a}$ Laboratoire de Physique Th\'eorique ENSLAPP
       \footnote{URA 1436 du CNRS, associ\'ee \`a l'Ecole Normale
       Sup\'erieure de Lyon et \`a l'Universit\'e de Savoie.}}
    \\
  {\em Chemin de Bellevue BP 110, F-74941 Annecy-le-Vieux Cedex, France}

  \vs{5}

  {\em $^{b}$ Dipartimento di Scienze Fisiche, Universit\`a di Napoli
       ``Federico II''}
     \\
  {\em and I.N.F.N., Sezione di Napoli, Italy}

  \vs{5}

  {\em $^{c}$ Dipartimento di Fisica Teorica, Universit\`a di Torino}
     \\
  {\em and I.N.F.N., Sezione di Torino, Italy}

\end{center}

\vfill

\begin{abstract}
We give a realization of quantum affine Lie algebra $\uqa$ in terms of
anyons defined on a two-dimensional lattice, the deformation parameter
$q$ being related to the statistical parameter $\nu$ of the anyons by
$q = e^{i\pi\nu}$.
In the limit of the deformation parameter going to one we recover the
Feingold-Frenkel \cite{FF} fermionic construction of undeformed affine
Lie algebra.
\end{abstract}

\vfill
\vfill

\rightline{ENSLAPP-AL-552/95}
\rightline{DSF-T-40/95}
\rightline{DFTT-60/95}
\rightline{hep-th/9511013}
\rightline{October 95}

\newpage
\pagestyle{plain}

\indent

\sect{Introduction}
\label{intro}

The connection between anyons and quantum groups was originally
discovered in \cite{LS} for the case of $Sl_q(2)$ and later extended
in \cite{CM} to $Sl_q(n)$ and in  \cite{FMS} to the other classical
deformed algebras.  For a review see \cite{FLS}.

Anyons are particles with {\it any} statistics
\cite{LM,W1} that interpolate
between bosons and fermions of which they can be considered, in some
sense, as a deformation (for reviews see for instance \cite{W2,L}).
Anyons exist only in two dimensions because in this
case the configuration space of collections of identical particles
has some special topological properties allowing arbitrary
statistics, which do not exist in three or more dimensions.
Anyons are deeply connected to the braid group of
which they are abelian representations, just like
bosons and fermions are abelian representations
of the permutation group. In fact, when one exchanges two identical
anyons, it is not enough to compare their final configuration
with the initial one,
but instead it is necessary to specify also the way in which
the two particles are exchanged, {\it i.e.} the way in which they
braid around each other.

In the present paper we will use creation and destruction anyonic
operators, equipped with a prescription about their braiding; such a
prescription also induces an ordering among the anyons themselves
\cite{LS}.

However we remark that anyons can consistently be defined also on
one dimensional chains; for simplicity, in this paper mainly we will
use anyons defined on  an infinite one-dimensional chain; at the end we
will extend our results to anyons defined on a two-dimensional lattice.

Here we extend the realization of classical deformed algebras to
the case of the deformed affine algebra $\uqa$. As the
fermionic construction of  undeformed  affine algebras, the anyonic
construction on a linear chain gives a value of the central charge
equal to one.
However at the end, see Sect. \ref{sect3}, we briefly discuss how to get
central charges equal to an integer positive number, considering
anyons in a two-dimensional lattice.
In the case of deformed affine algebras two kinds of explicit realization
are known and only for the case of affine $Sl_q(n)$: a construction in
terms of an (infinite) combination of the Cartan-Chevalley generators
of $Sl_q({\infty})$ build up with bilinears of deformed bosonic or
fermionic oscillators (\cite{Hay}) or a vertex operator construction
(\cite{FJ}).

The paper is organized as follows. In Sect. \ref{uqa}, after
recalling the fermionic realization of $\wA_{N-1}$, we present an
anyonic realization by the use of anyons defined on a one-dimensional
lattice $\ZZ$. In Sect. \ref{sect3} we explain
how more general representations of the quantum affine algebras can be
built by means of anyons defined on a two-dimensional lattice; then we
make some final remarks. The Appendix is devoted to a discussion of
the ordering of anyons on a chain and on a two-dimensional lattice.

\sect{Anyonic contruction of $\uqa$}
\label{uqa}

\indent

Consider the affine Lie algebra $\wA_{N-1}$ with generators denoted in the
Cartan-Weyl basis by $h_i^m$ (Cartan generators) and $e_\fra^m$ (root
generators) where $i=1,\dots,N-1$ and $m
\in \ZZ$. The root system of $A_{N-1}$ is given by
$\Delta = \{ \e_\rho - \e_\sigma, \ 1 \le \rho \ne \sigma \le N \}$,
the $\e_\rho$'s spanning the dual of the Cartan subalgebra of $gl_N$.
The generators satisfy the following commutation relations:
\subequations
\bea
&& \left[ \bigg. h_i^m,h_j^n \right] = \gamma m \delta_{m,-n} \delta_{ij}
\label{eqA30a} \\
&& \left[ \bigg. h_i^m,e_\fra^n \right] = \fra_i ~ e_\fra^{m+n}
\label{eqA30b} \\
&& \left[ \bigg. e_\fra^m,e_\frb^n \right] = \left\{ \begin{array}{ll}
\e(\fra,\frb) e_{\fra+\frb}^{m+n} & \qquad \smbox{if} \fra + \frb
\smbox{is a root} \\
\fra^i h_i^{m+n} + \gamma m \delta_{m,-n} & \qquad \smbox{if} \frb =
-\fra \\
0 & \qquad \smbox{otherwise}
\end{array} \right.
\label{eqA30c} \\
&& \left[ \bigg. h_i^m,\gamma \right] =
\left[ \bigg. e_\fra^m,\gamma \right] = 0
\label{eqA30d}
\ena
\endsubequations
where $\e(\fra,\frb) = \pm 1$ is the usual 2-cocycle and $\gamma$ is the
central charge of the algebra.
\\
One can also use the Serre-Chevalley presentation, in which the
algebra is described in terms of simple generators and relations
(the Serre relations), the only data being the entries
of the Cartan matrix $(a_{\a\b})$ of the algebra.
Let us denote the generators in the Serre-Chevalley basis by $h_\a$ and
$e_\a^\pm$ where $\a=0,1,\dots,N-1$. The commutation relations are:
\subequations
\bea
&& \left[ \bigg. h_\a,h_\b \right] = 0
\label{eqA31a} \\
&& \left[ \bigg. h_\a,e^\pm_\b \right] = \pm a_{\a\b} e^\pm_\b
\label{eqA31b} \\
&& \left[ \bigg. e^+_\a,e^-_\b \right] = \delta_{\a\b} \ h_\a
\label{eqA31c}
\ena
\endsubequations
together with the Serre relations
\be
\sum_{\ell=0}^{1-a_{\a\b}} (-1)^\ell
\left( \begin{array}{c} 1-a_{\a\b} \\ \ell \end{array} \right)
\left( e_\a^\pm \right)^{1-a_{\a\b}-\ell} e_\b^\pm
\left( e_\a^\pm \right)^{\ell} = 0
\label{eqA32}
\ee
The correspondence between the Serre-Chevalley and the Cartan-Weyl
bases is the following ($i=1,\dots,N-1$):
\be
\begin{array}{lll}
\bigg. h_i = h_i^0 \qquad &
e_i^+ = e_{\e_i-\e_{i+1}}^0 \qquad &
e_i^- = e_{\e_{i+1}-\e_i}^0 \\
\bigg. h_0 = \gamma - \sum_{j=1}^{N-1} h_j^0 \qquad &
e_0^+ = e_{\e_N-\e_1}^1 \qquad &
e_0^- = e_{\e_1-\e_N}^{-1}
\end{array}
\label{eqA33}
\ee

\vs{7}

Let us recall now the fermionic realization
of $\wA_{N-1}$ in terms of creation and annihilation
operators. Consider an infinite number of fermionic oscillators
$c_\rho(r), c_\rho^\dagger(r)$ with $\rho=1,\dots,N$ and $r \in \ZZ+1/2 =
\ZP$, which satisfy the anticommutation relations
\be
\left\{ \bigg. c_\rho(r),c_\sigma(s) \right\}
= \left\{ \bigg. c_\rho^\dagger(r),c_\sigma^\dagger(s) \right\} = 0
\bigbox{and} \left\{ \bigg. c_\rho^\dagger(r),c_\sigma(s) \right\}
= \delta_{\rho\sigma} \delta_{rs}
\label{eqA1}
\ee
the number operator being defined as usual by
$n_\rho(r) = c_\rho^\dagger(r)c_\rho(r)$.
\\
These oscillators are equipped with a normal ordered product such that
\be
:c_\rho^\dagger(r) c_\sigma(s): = \left\{ \begin{array}{ll}
c_\rho^\dagger(r) c_\sigma(s) & \smbox{if} s > 0 \\
- c_\sigma(s) c_\rho^\dagger(r) & \smbox{if} s < 0
\end{array} \right.
\label{eqA2}
\ee
and therefore:
\be
:n_\rho(r): = \left\{ \begin{array}{ll}
n_\rho(r)  & \smbox{if} r > 0 \\
n_\rho(r) - 1  & \smbox{if} r < 0
\end{array} \right.
\label{eqA2'}
\ee
Then a fermionic oscillator realization of the generators of
$\wA_{N-1}$ is given by
\subequations
\bea
&& h_i^m = \sum_{r \in \ZP} \left(:c_i^\dagger(r) c_i(r+m): -
:c_{i+1}^\dagger(r) c_{i+1}(r+m): \right) ~~~ (i=1,\dots,N-1),
\label{eqA34a} \\
&& e_{\e_\rho-\e_\sigma}^m = \sum_{r \in \ZP} c_\rho^\dagger(r)
c_\sigma(r+m),~~~~~~~~~  \rho \ne \sigma,~~~  1\le \rho,\sigma \le N
\label{eqA34b}
\ena
\endsubequations
Now, taking $e_{\e_\rho-\e_\sigma}^m$ and $e_{\e_\sigma-\e_\rho}^{-m}$
with let us say, $\rho>\sigma$ and $m \ge 0$, one can compute
\bea
\left[ \bigg. e_{\e_\rho-\e_\sigma}^m,e_{\e_\sigma-\e_\rho}^{-m} \right]
&=& \left[ \bigg. \sum_{r \in \ZP} c_\rho^\dagger(r) c_\sigma(r+m) ,
    \sum_{s \in \ZP} c_\sigma^\dagger(s) c_\rho(s-m) \right] \nonumber \\
&=& \sum_{s \in \ZP} \left( c_\rho^\dagger(s-m) c_\rho(s-m)
    - c_\sigma^\dagger(s) c_\sigma(s) \right) \nonumber \\
&=& \sum_{s \not\in [0,m]} \left(:n_\rho (s-m):
    - :n_\sigma (s): \right) + \sum_{s \in [0,m]}(: n_\rho (s-m):
    - :n_\sigma (s) : + 1 )  \nonumber \\
&=& - \left( \sum_{k=\sigma}^{\rho-1} h_k^0 \right) + m
\label{eqA35}
\ena
which proves that the central charge has value $\gamma=1$ in the
fermionic representation. Note that the value of the central charge is
intimately related to the definition of the normal ordered product
Eq. (\ref{eqA2'}); different definitions, like:
\subequations
\bea
&& :n_\rho(r): = n_\rho(r), ~~~~~~\forall r \in \ZP
\label{eqA3'a} \\
\hbox{or}
&& :n_\rho(r): = n_\rho(r)-1,~~~ \forall r \in \ZP,
\label{eqA3'b}
\ena
\endsubequations
would lead to $\gamma = 0$.
\\
It follows that a fermionic oscillator realization of the simple
generators of $\wA_{N-1}$ in the Serre-Chevalley basis is given by
$(\a=0,1,\dots,N-1)$
\subequations
\bea
&& h_\a = \sum_{r \in \ZP}h_\a(r)
\label{eqA3a} \\
&& e_\a^\pm = \sum_{r \in \ZP} e_\a^\pm (r)
\label{eqA3b}
\ena
\endsubequations
where $(i=1,\dots,N-1)$
\subequations
\bea
&& h_i(r) = n_i(r) - n_{i+1}(r)
       =  :n_i(r): - :n_{i+1}(r):
\label{eqS3a} \\
&& h_0(r) = n_N(r) - n_1(r+1)
       =  :n_N(r): - :n_1(r+1): + \delta_{r,-1/2}
\label{eqS3b} \\
&& e_i^+(r) =  c_i^\dagger(r) c_{i+1}(r), ~~~~ ~~~~
 e_0^+ =  c_N^\dagger(r) c_1(r+1)
\label{eqS3c} \\
&& e_i^-(r) =  c_{i+1}^\dagger(r) c_i(r), ~~~~ ~~~~
 e_0^-(r) =  c_1^\dagger(r+1) c_N(r)
\label{eqS3d}
\ena
\endsubequations
Inserting Eq. (\ref{eqS3b}) into Eq. (\ref{eqA3a}) and taking into
account that the sum over $r$ can be splitted into a sum of two
convergent series only after normal ordering, one again checks that
\be
h_0 = 1 + \sum_{r \in \ZP} :n_N(r): - \sum_{r \in \ZP} :n_1(r): ~
= 1 - \sum_{j=1}^{N-1} h_j \;,
\label{eqS4}
\ee
that is $\gamma = 1$.

\vs{7}

In order to obtain an anyonic realization of $\uqa$, we will replace
the fermionic oscillators by suitable anyons in the expressions of the
simple generators of $\uqa$ in the Serre-Chevalley basis. We introduce
therefore the following anyons defined on a one-dimensional lattice
$(r \in \ZP)$:
\be
a_\rho(r) = K_\rho(r) c_\rho(r)
\qquad 1 \le \rho \le N,
\label{eqA4}
\ee
and similar expressions for the conjugated operator $a_\rho^\dagger(r)$,
where  the disorder factor $K_\rho(r)$ is expressed as
\be
K_\rho(r) = \prod_{t \ne r} e^{-i \nu \Theta(r,t) :n_\rho(t):}
\label{eqA5}
\ee
$\nu \in [0,2)$ being the statistical parameter and $\Theta(r,t)$ a
suitably defined angle (see Refs. \cite{LS,FLS}).
\\
The anyonic ordering being the natural order of
the integers, in  Eq. (\ref{eqA5}) the anyonic angle (see Appendix and
Ref. \cite{FLS}) can be chosen as
\be
\begin{array}{ll}
\Theta(r,t) = +\pi/2 & \smbox{if} r > t \\
\Theta(r,t) = -\pi/2 & \smbox{if} r < t
\end{array}
\label{eqAbis}
\ee
By means of Eq. (\ref{eqAbis}) the disorder factor $K_\rho(r)$  can be
rewritten, using the sign function $\e(t) = |t|/t$ if
$t \ne 0$ and $\e(0) = 0$,
\be
K_\rho(r) = q^{ -\frac{1}{2} \sum_{t \in \ZP} \e(t-r) :n_\rho(t): }
\label{eqA7}
\ee
with $q = \exp(i\pi\nu)$.
\\
By a direct calculation, one can prove that the $a$-type operators
satisfy the following braiding relations for $r>s$:
\bea
&& a_\rho(r) a_\rho(s) + q^{-1} a_\rho(s) a_\rho(r) = 0 \nonumber \\
&& a^\dagger_\rho(r) a^\dagger_\rho(s) + q^{-1} a^\dagger_\rho(s)
a^\dagger_\rho(r) = 0 \nonumber \\
&& a^\dagger_\rho(r) a_\rho(s) + q\ a_\rho(s) a^\dagger_\rho(r) = 0
\nonumber \\
&& a_\rho(r) a^\dagger_\rho(s) + q\ a^\dagger_\rho(s) a_\rho(r) = 0
\label{eqA71}
\ena
and
\bea
&& a_\rho(r) a^\dagger_\rho(r) + a^\dagger_\rho(r) a_\rho(r) = 1
\nonumber\\
&& a_\rho(r)^2 =  a^\dagger_\rho(r)^2 = 0
\label{eqA72}
\ena
\\
which shows that the operators $a_{\rho}(r), a_{\rho}^\dagger(s)$ are
indeed anyonic oscillators with statistical parameter $\nu$.
Notice that the last equations (\ref{eqA72}) do not contain any
$q$-factor, the operators $a_{\rho}(r), a_{\rho}^\dagger(s)$  at the
same point thus satisfying standard anticommutation relations.
\\
Following Ref. \cite{LS}, we also introduce twiddled anyonic
oscillators:
\be
{\ta}_\rho(r) ={\tilde  K}_\rho(r) c_\rho(r)
\qquad 1 \le \rho \le N, \label{eqA4'}
\ee
where:
\be
{\tilde K}_\rho(r) = \prod_{t \ne r} e^{-i \nu {\tilde \Theta}(r,t)
:n_\rho(t):},
\label{eqA5'}
\ee
with the same statistical parameter $\nu$ and opposite braiding (and
ordering) prescription, corresponding to the choice
\be
\begin{array}{ll}
{\tilde \Theta}(r,t) = -\pi/2 & \smbox{if} r > t \\
{\tilde \Theta}(r,t) = +\pi/2 & \smbox{if} r < t
\end{array}
\label{eqA'bis}
\ee
and therefore to replace $q$ with $q^{-1}$ in Eqs. (\ref{eqA7}),
(\ref{eqA71}).

Let us add that anyons with different indices $\rho,\sigma$ obey the
standard anticommutation relations, for all values of $r,s \in \ZP$ and
for all values of the anyonic parameter $\nu$.
\\
Finally, the braiding relations between $a$-type and ${\ta}$-type
anyons are given by
\bea
&& \left\{ \bigg. {\ta}_\rho(r), a_\rho(s) \right\}
= \left\{ \bigg. {\ta}^\dagger_\rho(r),a^\dagger_\rho(s) \right\} = 0
\medbox{for all} r,s \in \ZP
\label{eqA76} \\
&& \left\{ \bigg. {\ta}^\dagger_\rho(r), a_\rho(s) \right\}
= \left\{ \bigg. {\ta}_\rho(r),a^\dagger_\rho(s) \right\} = 0
\medbox{for all} r \ne s \in \ZP
\label{eqA77}
\ena
and
\bea
&& \left\{ \bigg. {\ta}_\rho(r), a^\dagger_\rho(r) \right\} =
q^{ \ \sum_{t \in \ZP} \e(t-r) :n_\rho(t): } \nonumber \\
&& \left\{ \bigg. {\ta}^\dagger_\rho(r), a_\rho(r) \right\} =
q^{ - \sum_{t \in \ZP} \e(t-r) :n_\rho(t): }
\label{eqA78}
\ena
It is useful to remark that the following identities hold:
\be
a^\dagger_\rho(r) a_\rho(r) =  {\ta}^\dagger_\rho(r) {\ta}_\rho(r) =
n_\rho (r)
\label{eqS1}
\ee
and that normal ordering among anyons is defined exactly as in
Eq. (\ref{eqA2'}).

Everywhere in this paper the generators $E^+$ will be expressed in terms
of the oscillators $a$ and the generators $E^-$ in terms of the ${\ta}$
 as, in the deformed case, the $E^-$'s are not
simply equal to $(E^+)^\dagger$, but also the exchange
$q \leftrightarrow q^{-1}$ must be performed.

We can now build an anyonic realization of the quantum affine
Lie algebra $\uqa$ "anyonizing" Eqs. (\ref{eqA3a}-\ref{eqA3b}),
i.e. replacing  the
fermionic oscillators $c_\rho$ with the $a_\rho$'s in
Eq. (\ref{eqS3c}), with the $\ta_\rho$'s in Eq. (\ref{eqS3d}),
and, indifferently (see Eq. (\ref{eqS1})), with the
$a_\rho$'s or the $\ta_\rho$'s in the Cartan generators $h_i$
(see Eqs. (\ref{eqS3a}) and (\ref{eqS3b})).

\begin{theorem}
An anyonic realization of the simple generators of the quantum affine
Lie algebra $\uqa$ with central charge $\gamma = 1$
is given by ($\a = 0,1,\dots,N-1$)
\be
H_\a = \sum_{r \in \ZP} H_\a(r)
\qquad
E^\pm_\a = \sum_{r \in \ZP} E^\pm_\a(r)
\label{eqA8}
\ee
where $(1 \le i \le N-1)$
\subequations
\bea
&& H_i(r) = h_i(r) = :n_i(r): - :n_{i+1}(r): \nonumber \\
&& E^+_i(r) = a^\dagger_i(r) a_{i+1}(r) \nonumber \\
&& E^-_i(r) = {\ta}^\dagger_{i+1}(r) {\ta}_i(r)
\label{eqA10a} \\
&& \nonumber \\
&& H_0(r) = h_0(r) = :n_N(r): - :n_1(r+1): + \delta_{r,-1/2}
\nonumber \\
&& E^+_0(r) = q^{-\shalf \e(r+1/2)} \ a^\dagger_N(r) a_1(r+1)
\nonumber \\
&& E^-_0(r) = q^{-\shalf \e(r+1/2)} \ {\ta}^\dagger_1(r+1) {\ta}_N(r)
\label{eqA10b}
\ena
\endsubequations
\end{theorem}

\noindent
{\bf Proof}
\\
We must check that the realization Eqs. (\ref{eqA10a}-\ref{eqA10b})
indeed satisfy the quantum affine Lie algebra $\uqa$ in the
Serre-Chevalley basis, i.e.  for $\a,\b = 0,1,\dots,N-1$,
\subequations
\bea
&& \left[ \bigg. H_\a,H_\b \right] = 0 \label{eqA12a} \\
&& \left[ \bigg. H_\a,E^\pm_\b \right]
= \pm a_{\a\b} E^\pm_\b \label{eqA12b} \\
&& \left[ \bigg. E^+_\a,E^-_\b \right]
= \delta_{\a\b} ~ [H_\a]_{q_\a} \label{eqA12c}
\ena
\endsubequations
with the quantum Serre relations
\be
\sum_{\ell=0}^{1-a_{\a\b}} (-1)^\ell
\left[ \begin{array}{c} 1-a_{\a\b} \\ \ell \end{array} \right]_{q_\a}
\left( E_\a^\pm \right)^{1-a_{\a\b}-\ell} E_\b^\pm
\left( E_\a^\pm \right)^{\ell} = 0
\label{eqA13}
\ee
where the notations are the standard ones, i.e.
\be
[x]_q = \frac{q^x - q^{-x}}{q - q^{-1}} \;, \qquad
\left[ \begin{array}{c} m \\ n \end{array} \right]_q
= \frac{[m]_q!}{[n]_q![m-n]_q!} \;, \qquad [m]_q! = [1]_q \dots [m]_q
\label{eqA13bis}
\ee
$a_{\a\b}$ being the Cartan matrix of $\wA_{N-1}$ given by ($N \ge 3$)
\be
a_{\a\b} = \left(\begin{array}{ccccccc}
2 & -1 & 0 & \cdots & \cdots & 0 & -1 \\
-1 & 2 & -1 & \ddots &&& 0 \\
0 & -1 & 2 & \ddots & \ddots && \vdots \\
\vdots & \ddots & \ddots & \ddots & \ddots & \ddots & \vdots \\
\vdots && \ddots & \ddots & 2 & -1 & 0 \\
0 &&& \ddots & -1 & 2 & -1 \\
-1 & 0 & \cdots & \cdots & 0 & -1 & 2
\end{array}\right)
\label{eqA0}
\ee
and, for $N=2$, by
\be
a_{\a\b} = \left( \begin{array}{cc} 2 & -2 \\ -2 & 2 \end{array} \right)
\label{eqA0bis}
\ee
\\
Moreover in Eq. (\ref{eqA12c}) and Eq. (\ref{eqA13}) one has:
\be
q_\a = q^{(\fra_\a,\fra_\a)/2}
\label{eqS2}
\ee
$(\fra_\a,\fra_\a)$ being the length of the $\a^{th}$ simple root $\fra_\a$,
and therefore  $q_\a = q$ for all $\a = 0,1,\dots,N-1$.

\vs{7}

The proof of this theorem will be based on the observation that the Eqs.
(\ref{eqA12a}-\ref{eqA12c}) and (\ref{eqA13}), which define a generic
deformed affine
algebra $U_q({\widehat \cA})$, reduce to $U_q(\cA)$ if the dot
number $0$ is removed, and to another finite dimensional algebra
$U_q(\cA')$ if the dot number $0$ is kept and one or more other
suitable dots are removed.
In general, the relations defining $U_q({\widehat \cA})$ coincide with
the overlap of those defining $U_q(\cA)$ and $U_q(\cA')$;
therefore it will be enough to check that the equations defining
$U_q(\cA)$ and $U_q(\cA')$ are satisfied.

Thus our strategy will be the following:
\begin{itemize}
\item
step 1) -- prove that the generators written in the anyonic
representation (see Eq. (\ref{eqA8})) satisfy the
relations (\ref{eqA12a}-\ref{eqA12c}) and (\ref{eqA13}) for
$\a,\b \ne 0$.
This proof is performed using the technique outlined in Sect. 4 of
\cite{FLS}, that is going through the following steps:
\begin{itemize}
\item
step 1a) -- individuate a set of generators $\{h_i, e^\pm_i;\ i \ne 0\}$
of a low dimensional representation of $U_q(\cA)$;
\item
step 1b) -- show that, once that the anyonic oscillators $a$, $\ta$ are
expressed in terms of the fermionic ones, the set
$\{H_i, E^\pm_i;\ i \ne 0\}$ is related to the set
$\{h_i, e^\pm_i;\ i \ne 0\}$
by iterated coproduct in $U_q(\cA)$ and thus is a representation of
$U_q(\cA)$.
\end{itemize}
\item
step 2) -- Repeat the same procedure for $U_q(\cA')$, through the
following steps:
\begin{itemize}
\item
step 2a) -- individuate a set of generators
$\{h'_\a, e'^\pm_\a;\ \a \in I\}$ of a low dimensional representation
of $U_q(\cA')$; $I$ is a suitable subset of the extended Dynkin
graph of $U_q({\widehat \cA})$, with $0 \in I$;
\item
step 2b) -- show that, once that the anyonic oscillators $a$, $\ta$ are
expressed in terms of the fermionic ones, the set
$\{H_\a, E^\pm_\a;\ \a \in I\}$ is related to the set
$\{h'_\a, e'^\pm_\a;\ \a \in I\}$ by iterated coproduct in $U_q(\cA')$
and thus is a representation of $U_q(\cA')$.
\end{itemize}
\end{itemize}

In order to prove the present theorem, the step 1) is easily done
once one realizes that, inserting Eqs. (\ref{eqA4}) and
(\ref{eqA4'}), the expressions (\ref{eqA10a}-\ref{eqA10b}) simplify and
become
\be
E^\pm_\a(r) = e^\pm_\a(r) \ q^{\shalf \sum_{t \in \ZP} \e(t-r) h_\a(t)}
\label{eqA11}
\ee
where the generators
$h_\a(r) = H_\a(r) \vert_{q=1} = H_\a(r)$ and $e_\a^\pm(r) =
E_\a^\pm(r) \vert_{q=1}$,  coincide with those defined in Eqs.
(\ref{eqS3a}-\ref{eqS3d}), corresponding to the undeformed affine
algebra $\wA_{N-1}$.

In fact, for any fixed $r \in \ZZ +1/2$, the set $\{ h_i(r), e_i^\pm(r);
\ i = 1,\dots,N-1 \}$ is a representation of $A_{N-1}$ of spin 0
and 1/2, and thus also of $U_q(A_{N-1})$ \cite{FMS}; thanks
to Eqs. (\ref{eqA8}), (\ref{eqA11}), $H_\a,E^\pm_\a$ are the correct
coproduct in $U_q(A_{N-1})$.
Therefore  Eqs. (\ref{eqA12a}-\ref{eqA12c}), (\ref{eqA13}) surely
holds for $\a,\b \ne 0$.

For the step 2), we cut the extended Dynkin diagram by deleting a
dot $\mu \ne 0$; for any fixed $r \in \ZP$, the set
$\{ h_0(r), e_0^\pm(r), h_i(r), e_i^\pm(r), h_j(r+1), e_j^\pm(r+1) \}$,
where $\mu+1 \le i \le N-1$ and $1 \le j \le \mu-1$, is a spin 0, 1/2
representation of $A_{N-1}$, and thus a representation of
$U_q(A_{N-1})$ (step 2a)); again, thanks
to Eqs. (\ref{eqA8}), (\ref{eqA11}), $\{H_\a,E^\pm_\a; \a \ne \mu \}$
is the correct coproduct in $U_q(A_{N-1})$.
Therefore, Eqs. (\ref{eqA12a}-\ref{eqA12c})
and the quantum Serre relations (\ref{eqA13}) hold for any value of
the indices $\a, \b$, only excluding the couples $\a=0,\b=\mu$ or
$\a=\mu,\b=0$. However, one notes that these equations
obviously hold for $\vert \a - \b\vert \ge 2$ where $\a-\b$ is taken
modulo $N$. Therefore, for $N \ge 4$, we take, for instance,  $\mu=2$ and
for $N=3$ it is enough to repeat the argument twice, once for $\mu=1$
and once for $\mu=2$.

To complete  the proof, one has still to consider the case
of $U_q(\wA_1)$, whose Cartan matrix is reported in Eq. (\ref{eqA0bis}).
In this case the previous argument fails to prove Eq. (\ref{eqA12c})
and the quantum Serre relations (\ref{eqA13}), which take now the
following form
\subequations
\bea
&& \left( E_0^\pm \right)^3 E_1^\pm - \lambda \left( E_0^\pm
\right)^2 E_1^\pm E_0^\pm + \lambda E_0^\pm E_1^\pm \left(
E_0^\pm \right)^2 - E_1^\pm \left( E_0^\pm \right)^3 = 0
\label{eqA14a} \\
&& \left( E_1^\pm \right)^3 E_0^\pm - \lambda \left( E_1^\pm
\right)^2 E_0^\pm E_1^\pm + \lambda E_1^\pm E_0^\pm \left(
E_1^\pm \right)^2 - E_0^\pm \left( E_1^\pm \right)^3 = 0,
\label{eqA14b}
\ena
\endsubequations
where $\lambda = q^2+q^{-2}+1$.
Such equations can however be explicitly checked by using the braiding
properties of the anyonic oscillators $a$ and $\ta$.

The central charge is equal to 1 exactly for the same reasons of the
undeformed case. As the deformation of any affine Lie algebra does not
change its central charge, we will not repeat this argument in the
following.
\\
It is worthwhile to remark that the presence of a power of $q$ in front
of the anyonic oscillators in the expressions (\ref{eqA10b}) of
$E_0^{\pm}(r)$ (necessary in order to reproduce Eq. (\ref{eqA11}))
reflects the non vanishing of the central charge $\gamma$ and is related
to the definition Eq. (\ref{eqA2'}) of normal ordering; no power of $q$ in
front of the anyonic oscillators in Eq. (\ref{eqA10b}) would be needed if
 normal ordering were defined according to Eq. (\ref{eqA3'a}) or to
Eq. (\ref{eqA3'b}), corresponding to vanishing central charge.

\vs{7}

Let us remark that we could prove Theorem 1 by using the embedding
$U_q({\widehat A}_{N-1}) \subset U_q(A_{\infty})$ according to a theorem
due to Hayashi \cite{Hay}, that we report here in our notations:
\begin{theorem}
Let $k_i, f_i^\pm$ ($i \in \ZZ$) be the simple generators of
$U_q(A_{\infty})$, satisfying the following equations:
\subequations
\bea
&& \left[ \bigg. k_i,k_j \right] = 0
\label{eqH1a} \\
&& \left[ \bigg. k_i,f_j^\pm \right] = \pm a_{ij} f_j^\pm
\label{eqH1b} \\
&& \left[ \bigg. f_i^+,f_j^- \right] = \delta_{ij} ~ [k_i]_q
\label{eqH1c} \\
&& \sum_{\ell=0}^{1-a_{ij}} (-1)^\ell
\left[ \begin{array}{c} 1-a_{ij} \cr \ell \end{array} \right]_q
\left( f_i^\pm \right)^{1-a_{ij}-\ell} f_j^\pm
\left( f_i^\pm \right)^{\ell} = 0
\label{eqH2}
\ena
\endsubequations
with the infinite dimensional Cartan matrix
\be
a_{ij} = \left(\begin{array}{cccccccc}
\ddots & \ddots & \ddots & \ddots & \ddots &&& \cr
& 0 & -1 & 2 & -1 & 0 && \cr
&& 0 & -1 & 2 & -1 & 0 & \cr
&&& \ddots & \ddots & \ddots & \ddots & \ddots
\end{array}\right)
\label{eqH3}
\ee
Then
\be
H_\a = \sum_{\ell \in \ZZ} k_{\a+N\ell} \bigbox{and}
E_\a^\pm = \sum_{\ell \in \ZZ} f_{\a+N\ell}^\pm ~
q^{\shalf \sum_{m \in \ZZ} \e(m-\ell) k_{\a+Nm}}
\label{eqH4}
\ee
where $\a = 0,1,\dots,N-1$, give a representation of the simple
generators of $U_q({\widehat A}_{N-1})$.
\end{theorem}
By means of Eq. (\ref{eqA11}) one easily checks that the generators
$H_\a$ and $E_\a^\pm$ defined in Eq. (\ref{eqA8}) coincide with those
defined in Eq. (\ref{eqH4}), once that the identification
\bea
&& h_\a(\ell-1/2) = k_{\a+N\ell} \nonumber \\
&& e_\a^\pm(\ell-1/2) = f_{\a+N\ell}^\pm
\label{eqH5}
\ena
where $\ell \in \ZZ$ and $\a = 0,1,\dots,N-1$, is made.

\sect{More general representations  and two dimensional anyons}
\label{sect3}

In the previous section, we have built a representation of the
deformed affine Lie algebras $\uqa$ by means of anyons defined on an
infinite linear chain; as the corresponding fermionic representation,
it has central charge $\gamma = 1$.

Representations with vanishing central charge could be built in the
same way by using alternative normal ordering prescriptions Eq.
(\ref{eqA3'a}-\ref{eqA3'b}).

Representations with $\gamma = 1$ and $\gamma = 0$ can be combined
together by associating a representation to any horizontal line of a
two-dimensional square lattice, infinite in one direction, let us say
the horizontal one. So by $M$ copies of one-dimensional representations
with central charge equal to 1 one can get representations with the
value of the central charge equal to $M$. Note that by combining  one
representation with central charge equal to $M$ with a finite number of
(one-dimensional)
representations with vanishing value of the central charge one obtains
an inequivalent representation with the same value of the central
charge. We do not discuss here the problem of the irreducibility of
these representations. The extensions to two dimensional lattice
infinite in both directions can also be done, but it requires some
care in the definition in order to avoid convergence problems.

We will show now that the use of anyons defined on a two-dimensional
lattice naturally gives the coproduct of representations with the
correct powers of the deformation parameter $q$.
Each site of the two-dimensional square lattice is
labelled by a  vector $\vx = (x_1,x_2)$; the first
component $x_1 \in \ZP$ is the coordinate of
a site on the line $x_2 \in \ZZ$. As it is shown in the Appendix in
Eq. (\ref{eqApp5}), in a suitable gauge, the angle $\Theta(\vx,\vy)$
which enters into the definition of two-dimensional anyons can be
chosen in such a way that
\be
\Theta(\vx,\vy) = \left\{ \begin{array}{ll}
+\pi/2 & \qquad \smbox{if} x_2 > y_2 \cr
-\pi/2 & \qquad \smbox{if} x_2 < y_2
\end{array} \right.
\label{eqS01}
\ee
while when $\vx$ and $\vy$ lie on the same horizontal line,
that is $x_2=y_2$, the definitions of Sect. \ref{uqa} hold.
\\
Two-dimensional anyons still satisfy the braiding and anticommutations
relations expressed in the general form in Eqs.
(\ref{eqA72}-\ref{eqA78}).

Replacing in Sect. \ref{uqa} one-dimensional
anyons with two-dimensional ones, and making the corresponding
replacements in the sums over the sites, we easily get that
\be
H_\a = \sum_{\vx} H_\a(\vx) \qquad E^\pm_\a = \sum_{\vx} E^\pm_\a(\vx)
\label{eqS02}
\ee
where the sum over the vector of the lattice has to be read as the sum
on the (infinite) line $x_1$ and the sum over the (finite) line $x_2$,
that is
\bea
&& H_\a = \sum_{x_2} H_\a(line ~ x_2) \nonumber \\
&& E^\pm_\a = \sum_{x_2} q_\a^{\shalf \sum_{y_2} \e(y_2-x_2)
H_\a(line ~ y_2)} ~ E^\pm_\a(line ~ x_2)
\label{eqS03}
\ena
where $q_\a = q^{(\fra_\a,\fra_\a)/2}$ and $H_\a(line ~ x_2)$,
$E^\pm_\a(line ~ x_2)$ are the simple generators of the
$U_q(\widehat{\cA})$ defined in the previous section.
These equations show that the coproduct rules are fulfilled and the
set $\{ H_\a,E^\pm_\a\}$ gives a representation of the algebra
$U_q(\widehat{\cA})$ with central charge equal to the sum of the
central charges associated to each line of the two-dimensional lattice.

It is important to remark that, for sake of simplicity, we have proved
all the theorems of Sect. \ref{uqa} by exploiting the representation of
anyons in terms of fermions, the disorder operators giving the correct
coproducts; however, this is not necessary and purely "anyonic" proofs
of our Theorems can be given: it is easy to get convinced that the
braiding and anticommutations relations of
Eqs. (\ref{eqA72}-\ref{eqA78}) are the only relations necessary to
prove that the
simple generators built by means of anyons satisfy the commutations
relations and the Serre equations defining the deformed affine
algebras. We emphasize that anyons, defined by the braiding and
anticommutations relations of Eqs. (\ref{eqA72}-\ref{eqA78}), have
nothing to do with $q-oscillators$, that were even used to build
representations of deformed algebras.

In the previous sections we have discussed the case of $\vert q \vert
= 1$. The case of $q$ real can also be discussed and we refer to
\cite{LS} for the definition of anyons for generic $q$.

In conclusions we have presented a method to get representations (in
general reducible) of the affine untwisted quantum algebra $\uqa$ with
positive integer value of the central charge. The role of the
definition of the ordering of the anyons and of the normal ordering is
essential in the above construction. One can naturally ask if this
construction can also be generalized to the case of the other
affine untwisted quantum algebras ($B,C,D$ series) and to the twisted
affine algebras, or if a construction of a deformed Virasoro algebra can be
obtained  by means of anyons, or if anyons can also be used to realize
affine deformed Lie superalgebras. On the same line, one could think
that anyons are natural objects to use in statistical mechanical
models in which quantum groups arise.
We hope to address these issues in some future work.

\sect{Appendix}
\label{App}

A crucial role in the construction of anyons in terms of fermions
coupled to a Chern-Simons field, which endow them with a magnetic
flux, is given by the angle $\Theta(\vx,\vy)$ under which the point
$\vx$ is seen from the point $\vy$. As the angle $\Theta(\vx,\vy)$
appears in the exponent of the disorder operator $K(\vx)$
(\ref{eqA5}) multiplied by $i\nu$, $\nu$ being not integer, it must
be thoroughly defined without any ambiguity (see \cite{LS} and
references quoted therein).
On a two-dimensional lattice, we fix a base point $B$ and
associate to any point $\vx$ a path from $B$ to $\vx$ (see figure 1).

\begin{center}
\begin{picture}(50,50)
\thinlines
\multiput(-5,0)(0,10){5}{\line(1,0){50}}
\multiput(0,-5)(10,0){5}{\line(0,1){50}}
\put(0,0){\circle*{5}}
\put(40,40){\circle*{5}}
\put(80,0){\circle*{5}}
\put(5,5){\circle*{5}}
\put(-10,0){\makebox(0.4,0.6){$\vy$}}
\put(5,-10){\makebox(0.4,0.6){$\vy^*$}}
\put(40,50){\makebox(0.4,0.6){$\vx$}}
\put(80,-10){\makebox(0.4,0.6){$B$}}
\thicklines
\put(5,5){\line(1,1){35}}
\put(40,40){\line(1,0){10}}
\put(50,40){\line(0,-1){20}}
\put(50,20){\line(1,0){30}}
\put(80,20){\line(0,-1){20}}
\end{picture}

\vspace{5mm}

figure 1
\end{center}
The angle $\Theta(\vx,\vy)$ is defined as
\be
\Theta(\vx,\vy) = \widehat{B \vy^* \vx} + \theta_0
\label{eqApp1}
\ee
where $\theta_0$ is a constant and $\widehat{B \vy^* \vx}$ is the
angle under which the oriented path $B\vx$ is seen from a point
$\vy^*$ that will be eventually sent to $\vy$ inside one of the
four cells which have $\vy$ as vertex. The point $\vy^*$ is shifted from
the point $\vy$ by an arbitrarily small vector in order
to define unambiguously the angle $\widehat{B \vy^*
\vx}$ even if the path $B\vx$ passes through the site $\vy$.
The simplest choice is to fix $B$ as the infinity point of the
positive $x$-axis and to associate to each point $\vx$ the
horizontal straight line coming from $B$ (see figure 2).
\begin{center}
\begin{picture}(100,50)
\thinlines
\multiput(-5,0)(0,10){5}{\line(1,0){50}}
\multiput(0,-5)(10,0){5}{\line(0,1){50}}
\put(40,20){\circle*{5}}
\put(50,12){\makebox(0.4,0.6){$\vx$}}
\put(95,12){\makebox(0.4,0.6){$B$}}
\thicklines
\put(40,20){\line(1,0){60}}
\end{picture}

\vspace{5mm}

figure 2
\end{center}
Choosing $\vy^* = \vy + \e(\vu_x + \vu_y)$ with $\e \rightarrow
0^+$ and $\vu_x,\vu_y$ unit vectors of the axis $x$ and $y$, one
easily gets for two points on the same line ($x_2=y_2$)
\be
\begin{array}{ll}
\bigg. \widehat{B \vy^* \vx} = 0 & \qquad \smbox{if} x_1 > y_1
\cr
\bigg. \widehat{B \vy^* \vx} = -\pi & \qquad \smbox{if} x_1 < y_1
\label{eqApp1bis}
\end{array}
\ee
and therefore, fixing $\theta_0 = \pi/2$,
\be
\Theta(\vx,\vy) = \left\{ \begin{array}{ll}
+\pi/2 & \qquad \smbox{if} x_1 > y_1 \cr
-\pi/2 & \qquad \smbox{if} x_1 < y_1
\end{array} \right.
\label{eqApp1ter}
\ee
reproducing Eq. (\ref{eqAbis}).

For two generic points $\vx,\vy$ on the two-dimensional lattice, the
angle $\widehat{B \vy^* \vx}$ can get any value in the interval
$[-\pi,\pi)$. However, in the braiding relations, and therefore in all
calculations of this paper, the only quantities coming into play
are the differences $\Theta(\vx,\vy) - \Theta(\vy,\vx)$. It is easy
to check that, for any couple $\vx,\vy$, the definition of figure 2 gives
\be
\Theta(\vx,\vy) - \Theta(\vy,\vx) = \left\{ \begin{array}{ll}
+\pi & \begin{array}{l} x_2 > y_2 \cr x_1 > y_1 \smbox{if} x_2 = y_2
\end{array} \cr
&\cr
-\pi & \begin{array}{l} x_2 < y_2 \cr x_1 < y_1 \smbox{if} x_2 = y_2
\end{array}
\end{array} \right.
\label{eqApp2}
\ee
The definition of the angle $\Theta(\vx,\vy)$ induces an ordering
among the points of the lattice \cite{LS}. We say
\bea
&& \vx \succ \vy \bigbox{if} \Theta(\vx,\vy) - \Theta(\vy,\vx) = +\pi
\nonumber \\
&& \vx \prec \vy \bigbox{if} \Theta(\vx,\vy) - \Theta(\vy,\vx) = -\pi
\label{eqApp3}
\ena
This ordering coincides with the natural one for points on a
horizontal line:
\be
\vx \succ \vy ~~~ \Longleftrightarrow ~~~ x_1 > y_1 \smbox{for} x_2 = y_2
\label{eqApp3bis}
\ee
As it is possible \cite{FLS} to deform continuously the functions
$\Theta(\vx,\vy)$ with the only restriction of keeping fixed the
quantities $\Theta(\vx,\vy) - \Theta(\vy,\vx),$ we will choose
\be
\Theta(\vx,\vy) = \pm \frac{\pi}{2} \quad \Longleftrightarrow
\quad \vx \stackrel{\displaystyle{\succ}}{\prec} \vy
\label{eqApp4}
\ee
and therefore
\be
\Theta(\vx,\vy) = \left\{ \begin{array}{ll}
+\pi/2 & \qquad \smbox{if} x_2 > y_2 \cr
-\pi/2 & \qquad \smbox{if} x_2 < y_2
\end{array} \right.
\label{eqApp5}
\ee


\end{document}